\documentclass[runningheads]{svmult}

\usepackage{makeidx}   
\usepackage{graphicx}  
\usepackage{subeqnar}  
\usepackage{multicol}  
\usepackage{physprbb}  
\makeindex             

\begin{document}
\title*{A study of distant Ly$\alpha$ emitters in overdense regions}
\toctitle{A study of distant Ly$\alpha$ emitters in overdense regions}
%
%
\titlerunning{A study of distant Ly$\alpha$ emitters in overdense regions}
%
\author{Bram P. Venemans\inst{1}
\and Huub J. A. R\"ottgering\inst{1}
\and George K. Miley\inst{1}
}
\authorrunning{Bram Venemans et al.}
%
%
\institute{Leiden Observatory, Niels Bohrweg 2, 2333 CA Leiden, The Netherlands
}

\maketitle              

\begin{abstract}
Recently, we conducted a Very Large Telescope (VLT) large program to
search for forming clusters by looking for overdensities of Ly$\alpha$
emitters around high redshift radio galaxies. In total seven
proto-clusters were discovered, including a proto-cluster around the
radio galaxy MRC 0316--257 at $z \sim 3.13$. This structure has an
excess of Ly$\alpha$ emitters by a factor of 3 as compared to the
field, and the derived mass is 2--5 $\times 10^{14}$ M$_\odot$. The
Ly$\alpha$ emitters in the proto-cluster are on average bluer than
Lyman Break Galaxies (LBGs). Also, the galaxies are faint (sub $L_*$)
and small (half light radii $< 1.7$ kpc, which is smaller than the
average size of LBGs). This might indicate that, at least a fraction
of, Ly$\alpha$ emitters could be young ($\sim 10^6$ yr), nearly
dust-free, forming galaxies.
\end{abstract}

\section{VLT Large Program: overdense regions in the early Universe}
Forming clusters of galaxies (proto-clusters) could provide information on
the formation of large scale structure in the early Universe. Since a
substantial number of galaxies can be detected at the same redshift
and features in the spectral energy distribution can be studied
systematically by choosing the appropriate filters, they are also
ideal places to study galaxy evolution. By comparing proto-cluster
galaxies to field galaxies, changes in in galaxy properties as a
function of environment can be studied.

An important issue is where to search for these proto-clusters. During
the last decade evidence has accumulated that high redshift radio
galaxies (HzRGs, $z > 2$) are forming brightest cluster galaxies at
the centers of clusters or proto-clusters (for an overview, see e.g.\
\cite{ven03}). Supporting evidence includes: (i) HzRGs are amongst the
most massive ellipticals in the early Universe \cite{jar01,deb02},
(ii) they can have extreme radio rotation measures, indicative of
dense hot gas \cite{car97} and (iii) radio galaxies at $0.5 < z < 1.5$
lie in moderately rich clusters \cite{hil91,bes00,bes03}. A pilot
project conducted on the Very Large Telescope (VLT) to search for a
proto-cluster around radio galaxy PKS 1138--262 at $z = 2.16$ resulted
in the detection of 15 Ly$\alpha$ emitters within 1000 km s$^{-1}$ of
the central radio galaxy \cite{kur00,pen00}. This provided direct
evidence that distant luminous radio galaxies can be used as tracers
of proto-clusters.

We initiated a VLT large program to search for Ly$\alpha$ emitters
around HzRGs. To select the most likely progenitors of cD ellipticals,
the radio galaxies in this program satisfied the following criteria:
large radio luminosities, and bright optical and IR continua. To be
able to find Ly$\alpha$ emitters, the redshift of the radio galaxies
had to be suitable for Ly$\alpha$ imaging with the available VLT/FORS
narrowband filters. Two objects at high redshift (radio galaxies at
$z=4.1$ and one at $z=5.2$) were added and for which a custom
narrowband filter was purchased.

\begin{table}[t]
\caption{Overview of the observed radio galaxy fields. The Table shows
the name of the radio galaxy, its redshift, the number of candidate
Ly$\alpha$ emitters in the imaging ('IMG'), the number of
spectroscopically confirmed Ly$\alpha$ emitters, excluding the radio
galaxy ('SPC') and the velocity dispersion of the confirmed
cluster members.}
\begin{center}
\renewcommand{\arraystretch}{1.4}
\setlength\tabcolsep{5pt}
\begin{tabular}{llllll}
\hline\noalign{\smallskip}
Name of RG & $z$ & IMG & SPC & $\Delta v$ (km s$^{-1}$) & Notes \\
\noalign{\smallskip}
\hline
\noalign{\smallskip}
MRC 2048--272 & 2.06 & ~16 & 2 & N/A & High extinction \\
PKS 1138--262 & 2.16 & ~70 & 14 & $\sim 1000$ & See \cite{kur00,pen00,kur03} \\
MRC 0052--241 & 2.86 & ~73 & 37 & $\sim 900$ & {} \\
MRC 0943--242 & 2.92 & $\sim 70$ & 29 & $\sim 800$ & {} \\
MRC 0316--257 & 3.13 & ~85 & 31 & $\sim 625$ & {} \\
TN J2009--3040 & 3.15 & ~20 & 11 & $\sim 470$ & Radio loud quasar from
\cite{deb00} \\
TN J1338--1942 & 4.10 & ~50 & 32 & $\sim 350$ & See \cite{ven02} for
more details \\
TN J0924--2201 & 5.19 & $\sim 20$ & 6 & N/A & {} \\
\hline
\end{tabular}
\end{center}
\label{lptable}
\end{table}

In total, 20 nights on the VLT and 2 nights on Keck in 2001--2003 were
used to observe 7 radio galaxy fields with redshifts up to
5.2. In these fields, roughly 400 candidate Ly$\alpha$ emitters were
found, of which 162 were confirmed to be Ly$\alpha$ emitters near the
redshift of the radio galaxy (Table \ref{lptable}). All six fields
studied to sufficient depth turned out to be overdense in Ly$\alpha$
emitters, as compared to blank field surveys. The galaxy overdensities
in the narrowband filters are 3--5. The structures have sizes of
$>$ 3 Mpc and masses of 10$^{14-15}$ M$_\odot$
\cite{kur03,ven02}.

In Sect.\ \ref{0316}, we will describe the observations of one of the
radio galaxy fields, the field of 0316--257, in detail. In Sect.\
\ref{prop}, we will discuss the properties of the Ly$\alpha$ emitters.

\section{An overdensity around MRC 0316--257 at $z=3.13$}
\label{0316}

One of the targets in our program was the radio galaxy MRC 0316--257.
There were several additional reasons that this radio galaxy was
chosen. First of all, it was already known that the object had two
spectroscopically confirmed companions \cite{lef96}. Secondly, the
redshift of the radio galaxy ($z=3.13$) allows for an efficient search
for Lyman Break Galaxies (LBGs) and for [OIII] $\lambda$ 5007 \AA\
emitters using a $K$-band narrowband filter mounted on an infrared
camera.

\subsection{Imaging and spectroscopy}

The field surrounding 0316--257 was imaged in a narrowband for 6.5
hours, and 1.3 hours in both $V$ and $I$ with the VLT/FORS2 camera. As
part of an imaging program with the {\em Hubble Space Telescope}, the
field was observed for 3 orbits with the Advanced Camera for Surveys
(ACS) in the F814W filter. For each object detected in the narrowband
image, the line flux, continuum strength and continuum slope and their
uncertainties were calculated using the measured narrowband and $V$
and $I$ magnitudes. Subsequently, with the derived line flux and
continuum strength, the rest-frame equivalent width (EW$_0$) of the
objects was computed. Following \cite{ven02}, galaxies with a
rest-frame equivalent width EW$_0 >$ 15 \AA\ and a significance
$\Sigma \equiv$ EW$_0/\sigma_{\mathrm{EW}_0} >$ 3 were selected as
candidate Ly$\alpha$ emitters. 85 objects were marked as candidate
emitters. Follow-up FORS2 spectroscopy of 40 candidate Ly$\alpha$ emitters
revealed 33 emission line galaxies of which 2 turned out to be [OII]
$\lambda$ 3727 \AA\ emitters with $z \sim 0.35$ and 31 were confirmed
to be Ly$\alpha$ emitters at a redshift of $\sim$ 3.13.

\subsection{A proto-cluster at $z=3.13$}

The velocity distriburion of the 31 confirmed emitters has a FWHM of
1510 km s$^{-1}$, which is much smaller than the width of the filter
(FWHM $\sim 3500$ km s$^{-1}$), and the peak of the velocity
distribution is located within 200 km s$^{-1}$ of the redshift of the
radio galaxy. The volume density of Ly$\alpha$ emitters within our
narrowband filter is a factor 3.0 $\pm$ 0.8 higher as compared to the
volume density of field emitters (e.g.\ \cite{fyn03}). Because the
structure is not virialized (taking the velocity dispersion as a
typical velocity, it would take a galaxy at least 5 Gyr to cross the
structure, while the age of the Universe at $z=3.13$ is only 2.2 Gyr), the
virial theorem cannot be used to calculate the mass of the
proto-cluster. Therefore, following \cite{ste98}, we use the volume
occupied by the proto-cluster, the mean density of the Universe at
$z=3.13$, the measured galaxy overdensity and the bias parameter, the
computed mass is in the range 2--5 $\times 10^{14}$ M$_\odot$. Because
the structure does not seem to be bound in the FORS images, this mass
estimate is a lower limit.

\section{Properties of the Ly$\alpha$ emitters}
\label{prop}

\underline{\textbf{Sizes}} Including the radio galaxy, 19 of the 32
confirmed emitters were located within the ACS image. Two of these emitters
were not detected in the image to a depth of $I_{814} > 27.1$
arcsec$^{-2}$ (3$\sigma$).
The emitters that do have a counterpart in the ACS image show a range
of morphologies. Four of them, including the radio galaxy, appear to
consist of several small (half light radii $<$ 1.8 kpc) clumps, six
are single, resolved objects and the rest (seven) are unresolved,
single objects. The half light radii of the single objects range
between $< 0.8$ and $\sim 1.7$ kpc. Interestingly, this is smaller
than the average size of LBGs at a redshift of $\sim 3$ \cite{fer03},
which is 2.3 kpc.

\bigskip

\noindent
\underline{\textbf{Spectra}} The line full width half maximum of the
Ly$\alpha$ emission line ranges from 120 km s$^{-1}$ to 800 km
s$^{-1}$, with a median of $\sim 260$ km s$^{-1}$. One emitter has a
FWHM of almost 2500 km s$^{-1}$, and is likely to harbor an
AGN. Emitters with a high signal-to-noise spectrum show an asymmetric
line profile with an absorbed blue wing. In those cases, the lines
were fitted by a Gaussian emission line with one or more Voigt
absorption profiles. The inferred column densities of the absorbers
are in the range of $10^{14}$--$10^{15}$ cm$^{-2}$. Taking the half light
radius as measured in the VLT narrowband image as the radius of the
object, the amount of projected neutral HI is in the range of 5--400
M$_\odot$.

\begin{figure}[t]
\begin{center}
\includegraphics[width=.9\textwidth]{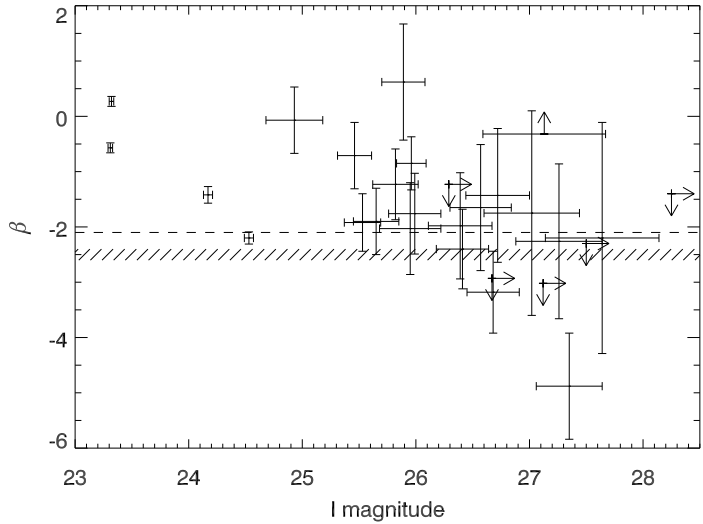}
\end{center}
\caption[]{I band magnitude plotted against UV continuum slope
  $\beta$. The dashed line represents the color of an unobscured,
  continuously star forming galaxy with an age $> 10^6$ yr
  \cite{lei99}. The area with parallel lines indicate the color of a
  young, unobscured starburst.}
\label{ivsbeta}
\end{figure}

\bigskip

\noindent
\underline{\textbf{UV colors}} Characterizing the spectral energy
distribution as $f_\lambda \propto \lambda^\beta$, the UV continuum
slope $\beta$ of the emitters was computed. In Fig.\ \ref{ivsbeta} the
slope $\beta$ is plotted against the $I$ magnitude of the confirmed
emitters. Excluding the radio galaxy and the emitter containing an
AGN, the median UV continuum slope of the confirmed emitters $\beta =
-1.65$.  Also, the color of a flux limited sample was determined. This
sample contained 23 candidate Ly$\alpha$ emitters with a Ly$\alpha$
flux $> 1.5 \times 10^{-17}$ erg s$^{-1}$ cm$^{-2}$, of which 20 are
confirmed. Again excluding the radio galaxy and AGN, the median slope
of the remaining emitters is $\beta = -1.65$, the same as the median
slope of the confirmed emitters. This is bluer than the average LBG
with Ly$\alpha$ in emission, which has a slope of $-1.09 \pm 0.05$
\cite{sha03}.

Models of galaxies with active star formation predict an UV continuum
slope around $\beta = -2.1$ for an unobscured, continuously star
forming galaxy with an age between a few $\times 10^6$ yr and over a
Gyr \cite{lei99}. 18 out of the 27 (67\%) confirmed Ly$\alpha$
emitters for which the slope are measured, have, within
their 1$\sigma$ errors, colors consistent with being an unobscured
star forming galaxy. Of those, 15 (56\%) have (within 1$\sigma$)
such blue colors ($\beta \sim -2.5$) that they could be star-forming
galaxies with very young ages ($\sim 10^6$ yr \cite{lei99}).

\begin{figure}[t]
\begin{center}
\includegraphics[width=.9\textwidth]{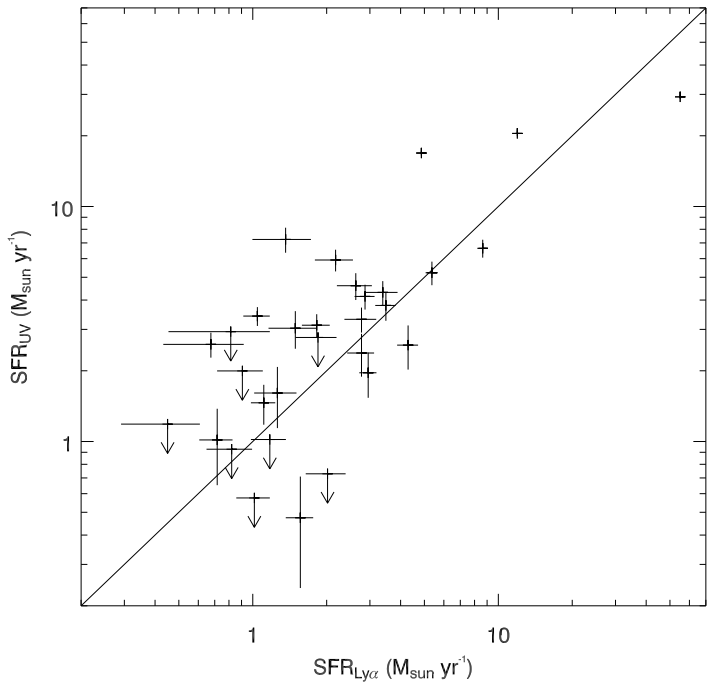}
\end{center}
\caption[]{Star formation rate as derived from the Ly$\alpha$ line
  luminosity, assuming Case B recombination, and the star formation
  rate derived from the UV continuum at 1500 \AA.}
\label{sfr}
\end{figure}

\bigskip

\noindent
\underline{\textbf{Star formation rates}} The star formation rates
(SFR) of the emitters were estimated in two ways. The first method was
to derive a H$\alpha$ luminosity from the Ly$\alpha$ luminosity
(assuming Case B recombination) and convert the H$\alpha$ luminosity
to a SFR following \cite{mad98}. The second method was to compute the
SFR from the UV luminosity density at a rest-frame wavelength of 1500
\AA\ \cite{mad98}. The Ly$\alpha$ line inferred SFR of the confirmed
emitters are 0.5--20 M$_\odot$ yr$^{-1}$, with an average of 2.6
M$_\odot$ yr$^{-1}$. In Fig.\ \ref{sfr} the SFR derived from the
Ly$\alpha$ flux is plotted against the UV continuum SFR rate. On
average, both methods to calculate the SFR give roughly the same
result.

\section{Summary and discussion}

Last year we finished a large program to search for proto-clusters
around HzRGs. Around all seven well studied radio galaxies
overdensities of Ly$\alpha$ emitters were found. This includes the
discovery of 31 confirmed Ly$\alpha$ emitters around the radio galaxy
MRC 0316--257 at $z \sim 3.13$. 

The Ly$\alpha$ emitters in this proto-cluster have blue continuum
slopes. Furthermore, the emitters are small and faint compared to
LBGs. Besides the radio galaxy, only one emitter is brighter than
$m_*$ at $z \sim 3$ \cite{ste98}, the other 30 confirmed emitters have
a continuum luminosity fainter than $L_*$. Given these properties, a
fraction of the Ly$\alpha$ emitters might be very young, forming
galaxies in their first starburst phase, without significant dust
absorption. Currently, we are investigating the other radio galaxy
fields to confirm this.

At $z \sim 3$ the star formation rate density (SFRD) derived from
observations of LBGs with $R \sim 27$ is 0.0184 $\pm$ 0.0034
M$_{\odot}$ yr$^{-1}$ Mpc$^{-3}$ \cite{ste99}. With the same magnitude
limit applied to the confirmed emitters, we find a SFRD of 0.0109
$\pm$ 0.0002 M$_{\odot}$ yr$^{-1}$ Mpc$^{-3}$ in the volume probed by
the narrowband filter. Because no correction has been made for
incompleteness and only confirmed emitters were considered, this is a
lower limit on the SFRD. Studies of LBGs showed that only 20--25\% of
the star forming, UV bright galaxies at $z\sim3$ have a Ly$\alpha$
line with an equivalent width $>$ 20 \AA\
\cite{ste00,sha03}. Correcting for this incompleteness, then the SFRD
around MRC 0316--257 is roughly 2.4--3.0 times higher than in the
field, in agreement with the overdensity computed from volume density
of the Ly$\alpha$ emitters.

%

\end{document}